# Analyzing Brain Activity During Learning Tasks with EEG and Machine Learning

**Ryan Cho[1], Mobasshira Zaman[2], Kyu Taek Cho[3], Jaejin Hwang[2*]**

[1]Illinois Mathematics and Science Academy, Aurora, IL
[2]Department of Industrial and Systems Engineering, Northern Illinois University, DeKalb, IL
[3]Mechanical Engineering Department, Northern Illinois University, DeKalb, IL.

Corresponding author's Email: jhwang3@niu.edu

**Abstract**

This study aimed to analyze brain activity during various STEM activities, exploring the feasibility of classifying between different tasks. EEG brain data from twenty subjects engaged in five cognitive tasks were collected and segmented into 4-second clips. Power spectral densities of brain frequency waves were then analyzed. Testing different k-intervals with XGBoost, Random Forest, and Bagging Classifier revealed that Random Forest performed best, achieving a testing accuracy of 91.07% at an interval size of two. When utilizing all four EEG channels, cognitive flexibility was most recognizable. Task-specific classification accuracy showed the right frontal lobe excelled in mathematical processing and planning, the left frontal lobe in cognitive flexibility and mental flexibility, and the left temporoparietal lobe in connections. Notably, numerous connections between frontal and temporoparietal lobes were observed during STEM activities. This study contributes to a deeper understanding of implementing machine learning in analyzing brain activity and sheds light on the brain's mechanisms.

Keywords: machine learning, electroencephalography, cognitive neuroscience, education

1. **Introduction**

Recent statistics have highlighted the rapid regression of students' math and reading skills to the worst levels (Mervosh, 2022). Amidst this massive decrease in education, this research explores if EEG sensors can be used as a teaching tool to help increase education rates. If teachers could monitor student learning patterns, they would have the ability to restructure lectures to be more engaging and adapt to the specific thinking processes that students prefer to use. Not to mention, it could also help teachers understand if problems are too difficult or easy to understand from students' Zone of Proximal Development (Vygotsky, 1978).

STEM (Science, Technology, Engineering, and Mathematics) learning encompasses a wide range of thinking methods for various types of activities. In this study, 'working memory', 'planning', 'arithmetic functioning', 'mental flexibility', and 'cognitive flexibility' were analyzed in detail. Working memory is crucial for tasks such as reasoning, verbal comprehension, and mathematical skills (Forsberg et al., 2021). Planning involves creating new learning strategies to approach a problem (Han et al., 2021). Arithmetic functioning involves understanding numerical information and performing calculations (Li, Wang, et al., 2020). Mental flexibility is the ability to generate original ideas and adapt to different situations (Elsayed & Abdo, 2022). Cognitive flexibility is the ability to switch between or think about multiple concepts simultaneously, making one look at multiple different angles of a problem (Cheng & Koszalka, 2016). While all these activities seem very similar, they create distinct patterns in the brain.

Each of these cognitive abilities represents different learning styles that students may prefer over others. These different preferences lead to different levels of success in the classroom depending on the learning style being used (M. Wilson, 2012). Recognizing these differences can help educators tailor their teaching strategies to meet the diverse needs of their students in the STEM fields (İlçin et al., 2018).

Traditional testing methods, such as quizzes, may not fully capture a student's real-time understanding and learning process. These assessments often focus on the product of learning rather than the process of thinking that leads to that answer (Li, Schoenfeld, et al., 2020). Furthermore, these tests are typically administered after a period of learning has occurred and may not provide timely feedback that can inform instruction and support a student's ongoing learning (Seo et al., 2021).

Noninvasive electroencephalogram (EEG) sensors can be used to provide a method of better understanding learning patterns in the classroom. By being able to actively monitor student brain waves during class, patterns of STEM learning methods can be identified, helping to formulate the best method of learning for a student. EEG sensors and machine learning in a learning setting can allow teachers to adapt their lectures based on what the majority of the students like to use to learn (İlçin et al., 2018). This is a much better method than just relying on teacher observations on student learning, especially due to the lack of awareness teachers have towards student performances other than the use of tests (Wadmare et al., 2022).

Before conducting this research, a comprehensive review of existing EEG studies on learning was created. Several research studies in the past have utilized EEG sensors and machine learning algorithms as a form of feedback in learning processes. Qu et. al analyzed passive and active learning tasks for STEM learning using four EEG sensors and collected data from twelve subjects. This research paper applied the Fast Fourier Transform to get power spectral density estimations and classified between passive and active states for both coding and math scenarios at an accuracy of over 90% for each task using a combination of boosting and bagging algorithms (Qu et al., 2018). Fitzgibbon et. al did a very analysis of eight cognitive tasks in 20 subjects wearing 64 EEG channels on their heads and performed statistical analysis on the differences in low amplitude gamma oscillations (Fitzgibbon et al., 2004). The study found that mental activity can increase gamma activity, which was recorded using EEG during various tasks. The increase was most noticeable during tasks involving expectancy, learning, reading, and subtraction, with 2–5 fold increases in gamma power at posterior and central scalp sites. Wilson et. al also used EEG data to classify fourteen different cognitive tasks performed by seven different subjects from sixty-second brain data recordings, achieving an 86% accuracy rate (G. F. Wilson & Fisher, 1995). This was accomplished with Stepwise Discriminant Analysis (SWDA) and Principal Components Analysis (PCA) to identify informative EEG frequency bands. Amin et al analyzed EEG data from 30 children with ADHD and 30 healthy children while performing a cognitive task. After extracting various signal features, the results showed that the MLP neural network achieved an accuracy of 92.28% and 93.65% using the minimum Redundancy Maximum Relevance and double input symmetrical relevance methods respectively (Amin et al., 2015).

The objective of this research is to identify important locations of the brain during STEM learning and contribute to the understanding of classifying STEM learning tasks through the analysis of brain signals. This leads to the development of EEG learning markers that can serve as an innovative tool for assessing students' cognitive states during STEM education. This study analyzed the frontal and temporal lobes of the brain through four EEG sensors and separated each signal into four-second signal windows. From each window, key features from the Fast Fourier Transform's periodogram are extracted. These features are the power spectral densities of the brain waves: delta, theta, alpha, beta, and gamma. These waves stand to be a major feature in the analysis of human behavior through EEG signals (Abhang et al., 2016). The machine learning algorithms of random forest, XGBoost, and bagging classifier are all trained equally using 100 trees each on these extracted features, quantifying relationships between STEM learning and brain signals through the evaluation metric results of each algorithm (Saeidi et al., 2021). During the training of these algorithms, a variation of the k-fold cross-validation will be used to avoid data biases (Qu et al., 2018). These machine learning algorithms can provide better information on which region of the brain is most important during various STEM learning tasks and provide further information on the effectiveness of using EEG sensors to better understand student learning methods. It was hypothesized that classification of these STEM learning tasks even with a decrease in the number of EEG channels being used compared to previous studies can still lead to successful accuracy results.

## 2. Methods

### 2.1 Subjects

The brain data from 20 data collection subjects were used. Their mean and standard deviation for age is 24.1 ± 2.85. There were 18 male and 2 female subjects, 13 master's students and 7 bachelor's students, and 13 industrial engineering, 6 mechanical engineering, and 1 computer science major.

### 2.2 Data Collection

To analyze five various cognitive effects, students performed five different STEM tasks during their EEG data collection. Each subject wore the Bluetooth-enabled Muse® headsets due to the benefits of cost, easy handling, and operation. The Muse headset consists of four dry EEG sensors located at the AF7, AF8, TP9, and TP10 channels. While each subject performs the STEM tasks, data can be recorded onto the Muse Monitor app through a Bluetooth connection wirelessly onto a mobile phone and saved into a CSV file. The total data collection duration for each participant will be less than 2 hours. To be able to induce these different cognitive tasks, the Psychology Experiment Building Language (PEBL) software to run the STEM tasks (*PEBL: The Psychology Experiment Building Language*, n.d.). The STEM tasks being focused on for this study are working memory, planning, arithmetical functioning, mental flexibility, and cognitive flexibility. The PEBL software works so that as users immerse themselves in these STEM processes, the tasks increase in difficulty for every successful answer. The different types of games used in the PEBL software can be seen in Table 1 and Figure 1.

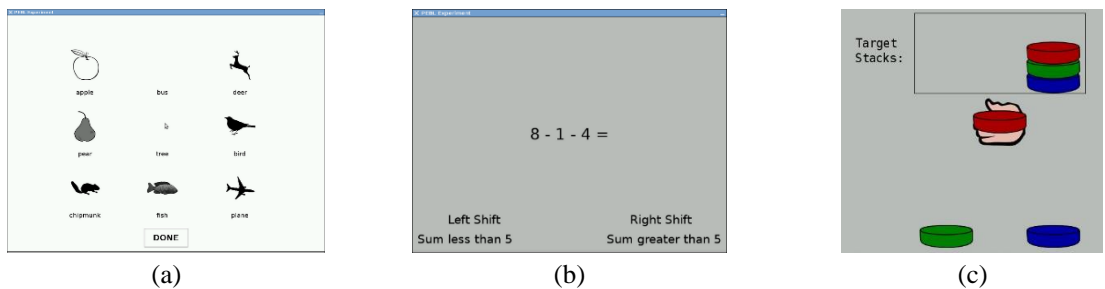

(a)      (b)      (c)

**Figure 1** Examples of neurocognitive tasks: (a) Memory Span, (b) Math Processing Task, and (c) Tower of London Task

Table 1. PEBL Activities (Mueller & Piper, 2014)

| STEM Learning | Task Name | Task Description |
|---|---|---|
| Working Memory | Memory Span (MSPAN) | Memorize and match up to 9 pictures that appear in sequence. |
| Arithmetical Functioning | Math Processing (MathProc) | Answer yes/no to whether the correct answer to a simple arithmetic problem is less than or greater than a certain number. |
| Cognitive flexibility | Wisconsin Card Sort (BCST) | Match stimulus cards based on changing rules which can be inferred. |
| Mental flexibility | Salthouse Connections (Connections) | Alternate alphabets and numbers and connect them in order. |
| Planning | Tower of London (TOL) | Build a tower of the target arrangement with minimal movement from the currently stacked tower |

### 2.3 Data Processing

After the data is recorded, muscular artifacts such as eye blinks and jaw clenches need to be removed either through methods such as manual removal, wavelet denoising, and bandpass filters (Jiang et al., 2019). However, the Muse Monitor mobile app automatically marks down areas where eye blinks and jaw clenches are found, allowing this study to efficiently sort through the data for clean pieces of data without any long manual inspections. After filtering through the time series data, four-second clips are extracted from each time series with each clip having a 90% temporal overlap with one another. This overlap between windows helps increase the temporal analysis of the nonstationary signals (Gevins et al., 1999).

After that, a commonly used method to better understand brain activity includes applying the Fast Fourier Transform (FFT) onto a time series (Singh et al., 2015). The application of the FFT, equation 1, allows for the analysis of the power spectral densities of different oscillations represented in the time series. Brain waves measured in the Muse have oscillating voltages bearing amplitudes from microvolts to millivolts, and they can be categorized into five bands depending on their frequency ranges, and each band exhibits states of the brain as shown in Table 2. Thus, the brain states can be understood by monitoring the brain signals of people in various tasks or actions.

Table 2. Frequency Bands (Abhang et al., 2016).

| Frequency band | Frequency | Brain-states |
|---|---|---|
| Gamma | > 35 Hz | Motor Functions, higher mental |
| Alpha | 12-35 Hz | Normal waking state, concentration, focus, five physical senses, integrated |
| Beta | 8-12 Hz | Relaxed, light meditation, creative, super learning, conscious |
| Theta | 4-8 Hz | Light sleep, deep meditation, creative, recall, fantasy |
| Delta | 0.5-4 Hz | Deep, dreamless sleep, non-REM sleep, unconscious |

Fast Fourier Transform (Owen, 2007):

$$X(k) = \frac{1}{N}\sum_{n=0}^{N-1} x(n) \cdot e^{-j\frac{2\pi}{N}kn} \tag{1}$$

Where N denotes the total length of signal, n represents each index within the signal, and x represents a time series list of all the data points of the signal.

*2.4 Machine Learning Algorithms*

The application of machine learning algorithms is crucial in deriving meaningful insights from the extracted features and their relevance to STEM learning task labels. Here, we delve deeper into the three decision tree ensemble algorithms used: Bagging Classifier, Random Forest, and XGBoost. A Bagging Classifier is a bootstrap aggregation technique that constructs an ensemble by resampling the training data and averaging the predictions (Mohammed & Kora, 2023). This method works by creating multiple subsets of the original data, training a model on each subset, and combining the results. The main advantage of bagging is that it reduces variance without increasing bias. This means that while individual predictions may be highly sensitive to noise in the training data, the average prediction of a large number of classifiers is not, as long as the classifiers are independent. Random Forest is an extension of bagging that adds an element of random feature selection, enhancing diversity among the trees (Biau & Scornet, 2016). It operates by constructing a multitude of decision trees at training time and outputting the class that is the mode of the classes (classification) or mean prediction (regression) of the individual trees. Random forests correct for decision trees' habit of overfitting to their training set. The randomness injected into the forest creation process leads to a wide diversity that generally results in a better model. XGBoost, short for Extreme Gradient Boosting, employs a gradient-boosting framework (Chen & Guestrin, 2016). It iteratively refines the ensemble by training new trees to correct the errors of previous ones. Unlike bagging and random forests, which train each tree independently, boosting trains each tree in a sequence. Each new tree is specifically designed to correct the mistakes made by the previous sequence of trees in the ensemble. The trees are added one at a time, and existing trees in the model are not changed. A gradient descent procedure is used to minimize the loss when adding trees. In this comparison, an equal configuration of 100 decision trees was employed for all classification tasks. This stable number of decision trees ensures there won't be any biases where one algorithm may perform better or worse than another merely because of the number of decision trees being used.

*2.5 Evaluation Methods*

To facilitate training and testing, a modified version of cross-validation is employed. While traditional cross-validation is highly effective during training, it comes with a significant computational burden. In this approach, the dataset is organized in chronological order. Subsequently, every kth interval of samples is selected and allocated to a training subset, leaving the remaining data for testing (Qu et al., 2018). This technique allows for unbiased evaluation of machine learning algorithms without incurring extensive computational costs. Moreover, it effectively gauges an algorithm's performance on previously unseen data by testing various interval sizes. This study systematically experimented with interval sizes ranging from 4096 to 2. The rationale behind these interval sizes is to thoroughly evaluate the machine learning algorithm's performance across varying data sizes, ranging from very small to progressively larger increments. This approach facilitates a more effective detection of potential biases within the algorithm.

Post-training, the algorithms are evaluated using four key metrics—overall accuracy, F1 score, precision, and recall—computed using Equation 1, on the designated testing dataset. The F1 score, a measure of the harmonic mean between precision and recall, assesses the algorithm's balance between positive predictions and actual positives. Precision quantifies the proportion of true positives among all predicted positives, while recall evaluates the ratio of actual positives correctly identified by the algorithm. All the equations shown in Table 3 are based on the True Positives (TP), False Positives (FP), False Negatives (FN), and True Negatives (TN) scores.

Table 3 Evaluation Metrics (Kumari et al., 2022).

| Metrics | Formula |
| --- | --- |
| Testing accuracy | $\dfrac{TP + TN}{TP + FP + TN + FN}$ |
| Precision | $\dfrac{TP}{TP + FP}$ |
| Recall | $\dfrac{TP}{TP + FN}$ |
| F1 Score | $\dfrac{2}{\dfrac{TP + FN}{TP} + \dfrac{TP + FP}{TP}}$ |

TP: True Positives, FP: False Positives, FN: False Negatives, and TN: True Negatives

3. **Results**

For the results, several different case scenarios were tested. The machine learning models were compared, the overall distribution of each cognitive task's power spectral densities was analyzed, the importance of each brain lobe location was analyzed, and then correlations between each brain region and their task were made.

*3.1 Algorithm Accuracies*

After training and testing all of the machine learning algorithms, the Random Forest has been found to perform the best at an accuracy of 91.07 %, as seen in Figure 2. It is also seen that at the higher intervals, the accuracy was significantly lower, but as more data is compiled for training, the accuracy increases. In Figure 1, it is noticed that regardless of the algorithm, as the interval size decreases, the accuracy grows like an exponential function. Random Forest performed an accuracy greater than 90% on an interval size of 2, which is the equivalent of a training size of 50%, showing very low signs of biases.

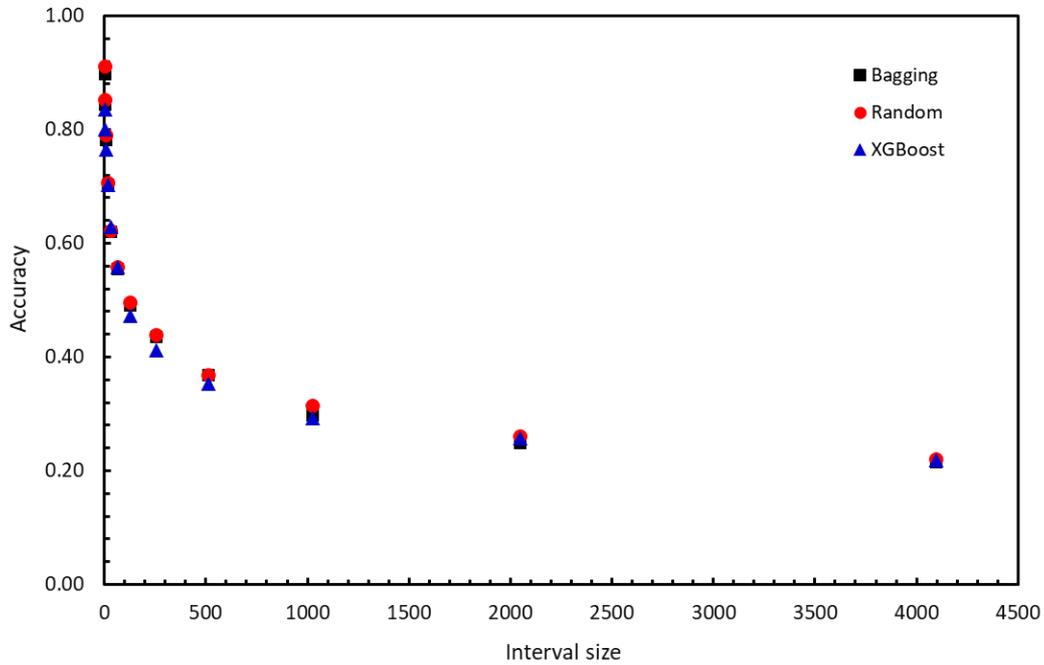

**Figure 2** Machine Learning Algorithm Evaluations on four EEGs. The Random Forest algorithm is seen to perform the best in classification accuracy most of the time for all the interval sizes tested

It is also seen from Table 4; much more detailed overview of each evaluation metric results for each interval. With the accuracy results being similar to the other metric scores, no major biases can be found in the algorithms.

Table 4. Comprehensive Overview of Algorithm Evaluations.

| Interval | Classifier | Accuracy | F1 Score | Precision | Recall |
|---|---|---|---|---|---|
| 4096 | Bagging Classifier | 21.44% | 20.96% | 23.98% | 21.95% |
| 4096 | Random Forest | 22.03% | 21.15% | 24.69% | 22.22% |
| 4096 | XGBoost Classifier | 21.83% | 21.71% | 22.70% | 22.16% |
| 2048 | Bagging Classifier | 24.91% | 24.92% | 25.36% | 24.95% |
| 2048 | Random Forest | 26.01% | 25.86% | 26.44% | 25.91% |
| 2048 | XGBoost Classifier | 25.77% | 25.46% | 26.35% | 25.87% |
| 1024 | Bagging Classifier | 29.77% | 29.57% | 29.66% | 29.64% |
| 1024 | Random Forest | 31.41% | 30.95% | 32.08% | 31.28% |
| 1024 | XGBoost Classifier | 29.29% | 29.27% | 29.29% | 29.33% |
| 512 | Bagging Classifier | 36.75% | 36.57% | 37.87% | 36.55% |
| 512 | Random Forest | 36.85% | 36.50% | 37.99% | 36.47% |
| 512 | XGBoost Classifier | 35.26% | 34.95% | 35.24% | 34.88% |
| 256 | Bagging Classifier | 43.53% | 43.04% | 43.84% | 42.96% |
| 256 | Random Forest | 43.87% | 43.37% | 43.99% | 43.27% |
| 256 | XGBoost Classifier | 41.20% | 40.58% | 41.02% | 40.56% |
| 128 | Bagging Classifier | 49.07% | 48.82% | 48.87% | 48.81% |
| 128 | Random Forest | 49.65% | 49.23% | 49.45% | 49.15% |
| 128 | XGBoost Classifier | 47.19% | 46.88% | 46.93% | 46.87% |
| 64 | Bagging Classifier | 55.48% | 55.02% | 55.10% | 55.00% |
| 64 | Random Forest | 55.81% | 55.38% | 55.52% | 55.32% |
| 64 | XGBoost Classifier | 55.82% | 55.47% | 55.63% | 55.41% |
| 32 | Bagging Classifier | 62.01% | 61.57% | 61.80% | 61.47% |
| 32 | Random Forest | 62.18% | 61.70% | 61.98% | 61.61% |
| 32 | XGBoost Classifier | 62.80% | 62.48% | 62.72% | 62.38% |
| 16 | Bagging Classifier | 70.31% | 69.93% | 70.16% | 69.82% |
| 16 | Random Forest | 70.61% | 70.23% | 70.47% | 70.11% |
| 16 | XGBoost Classifier | 70.22% | 69.90% | 70.13% | 69.79% |
| 8 | Bagging Classifier | 78.22% | 77.97% | 78.12% | 77.89% |
| 8 | Random Forest | 79.03% | 78.78% | 78.95% | 78.68% |
| 8 | XGBoost Classifier | 76.54% | 76.30% | 76.45% | 76.22% |
| 4 | Bagging Classifier | 84.35% | 84.21% | 84.32% | 84.14% |
| 4 | Random Forest | 85.19% | 85.07% | 85.15% | 85.01% |
| 4 | XGBoost Classifier | 80.04% | 79.85% | 79.95% | 79.79% |
| 2 | Bagging Classifier | 89.80% | 89.75% | 89.82% | 89.71% |
| 2 | Random Forest | 91.07% | 91.05% | 91.09% | 91.01% |
| 2 | XGBoost Classifier | 83.53% | 83.41% | 83.47% | 83.37% |

Then, the algorithms were also tested on their abilities to perform each separate task to notice any alarming patterns or irregularities to take note of as seen in Figure 3. It is seen from all the graphs in Figure 3 that the Random Forest performed the highest on all classification labels. The next major takeaway from this analysis was that the BCST activity had the highest-level accuracy of classification for all algorithms at interval size 2, meaning that it was the most easily recognizable.

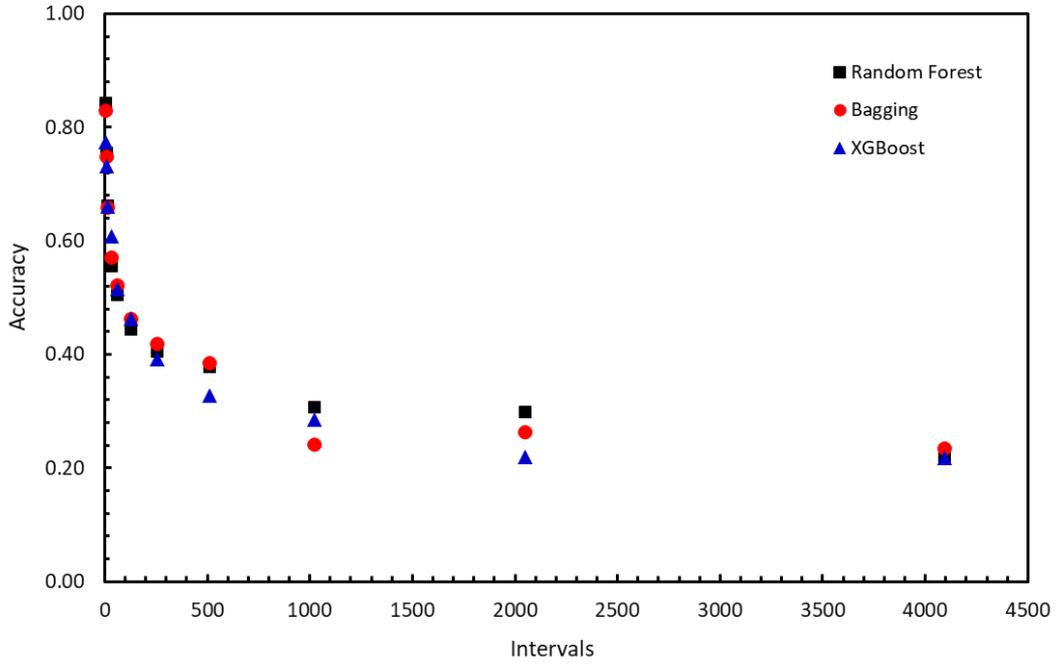

(a) MSPAN Classification Scores

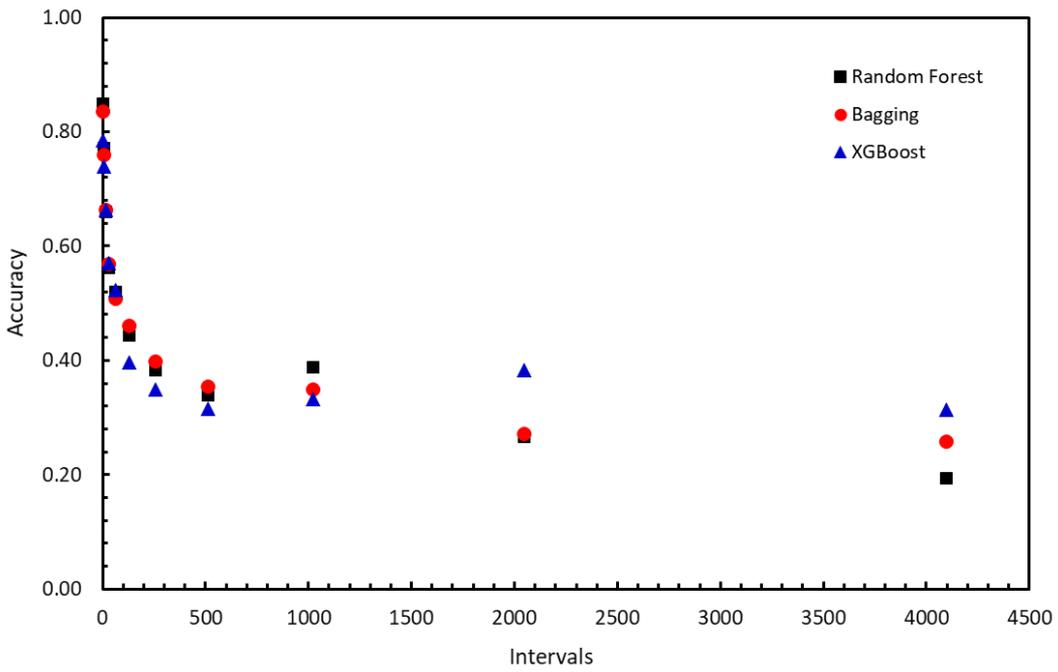

(b) TOL Classification Scores

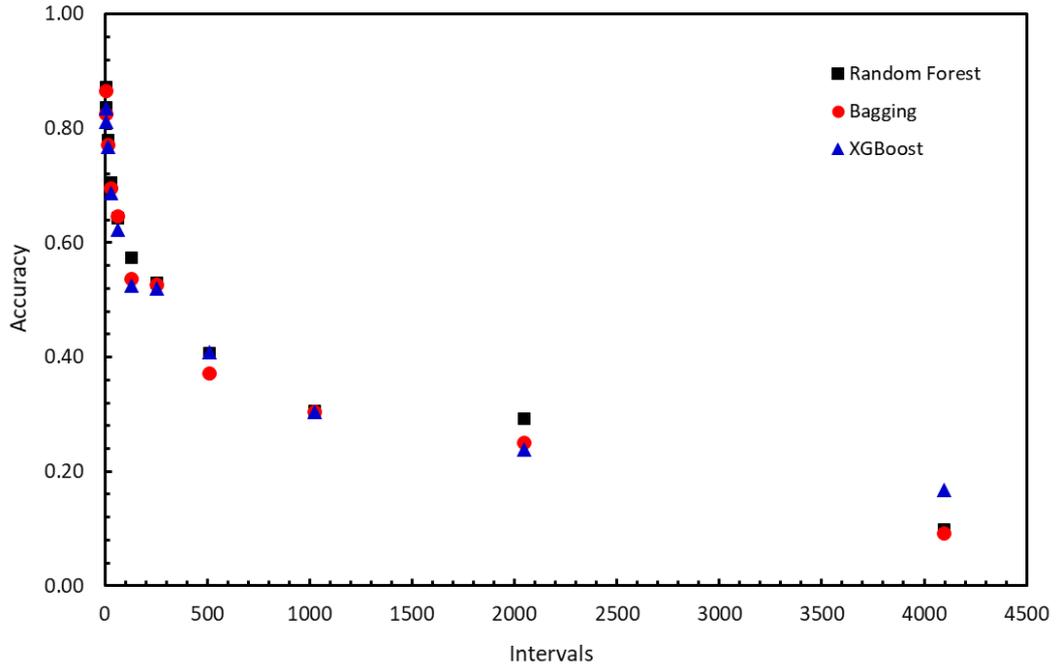

(C) MathProc Classification Scores

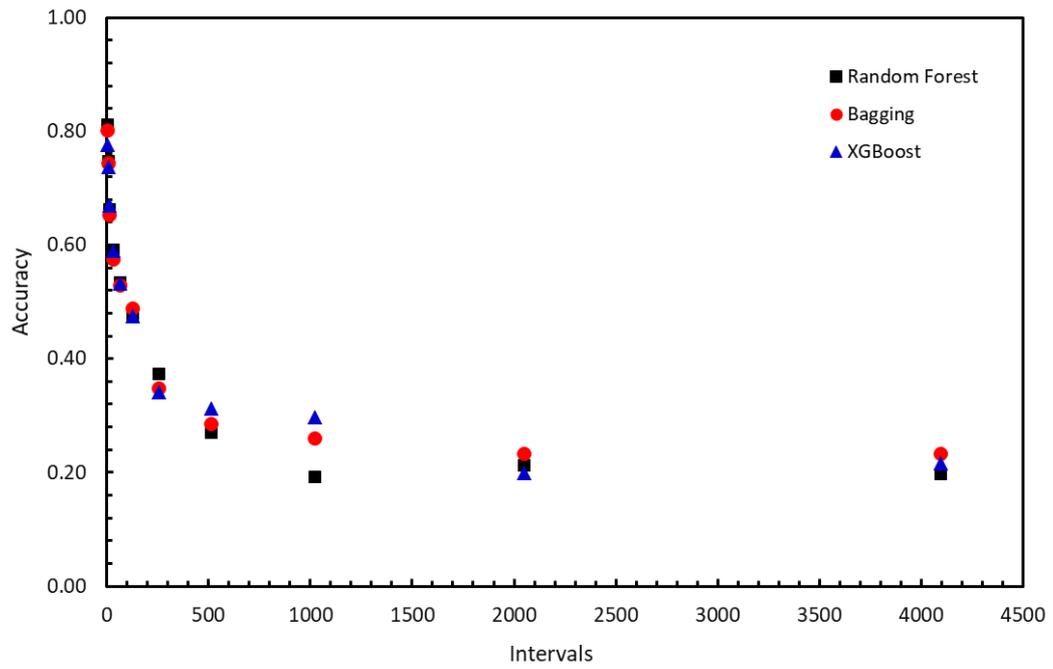

(d) Connection Classification Scores

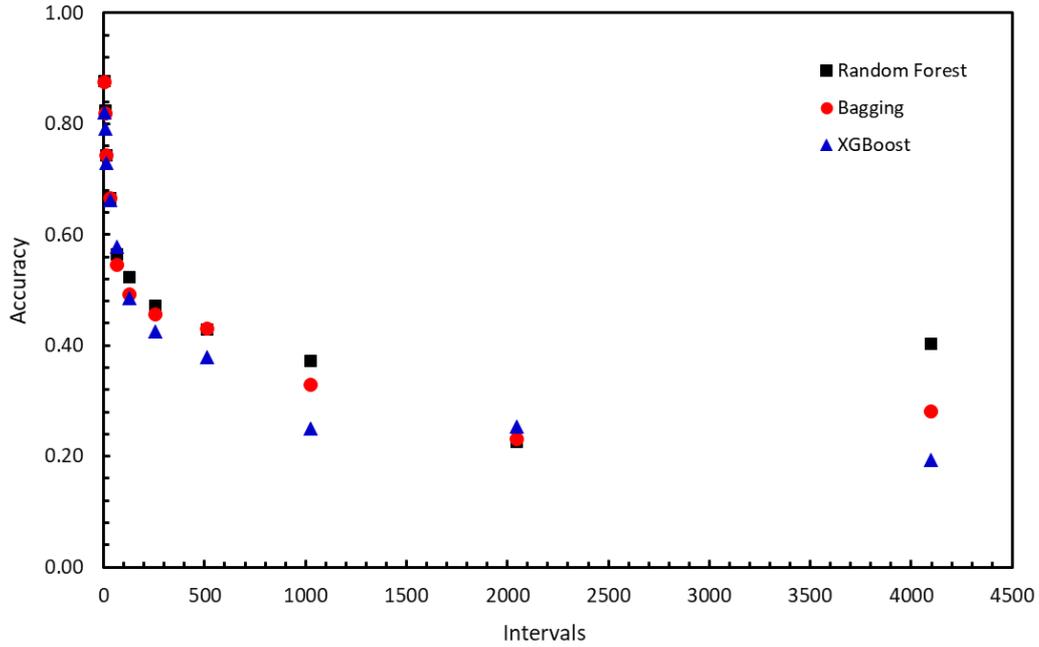

(e) BCST Classification Scores

**Figure 3** Classification Label Evaluations on Four EEGs. For all the different classification labels, the Random Forest performed at the highest accuracy. It is also noticed that the BCST activity has the highest classification accuracy compared to all the other activities

*3.2 Comparison of Brain Lobe Processes*

After evaluating this, the accuracy of relying on only one EEG channel was also tested using an interval size of 2 and the Random Forest algorithm, since they performed the best. This was conducted to see which EEG channel provides the most important and accurate information during classification. As seen in Figure 4, the AF8 EEG performed the best at an accuracy of around 54.68% with similar F1 score, precision, and recall scores. After the AF8, the AF7, TP9, and then the TP10 performed the best.

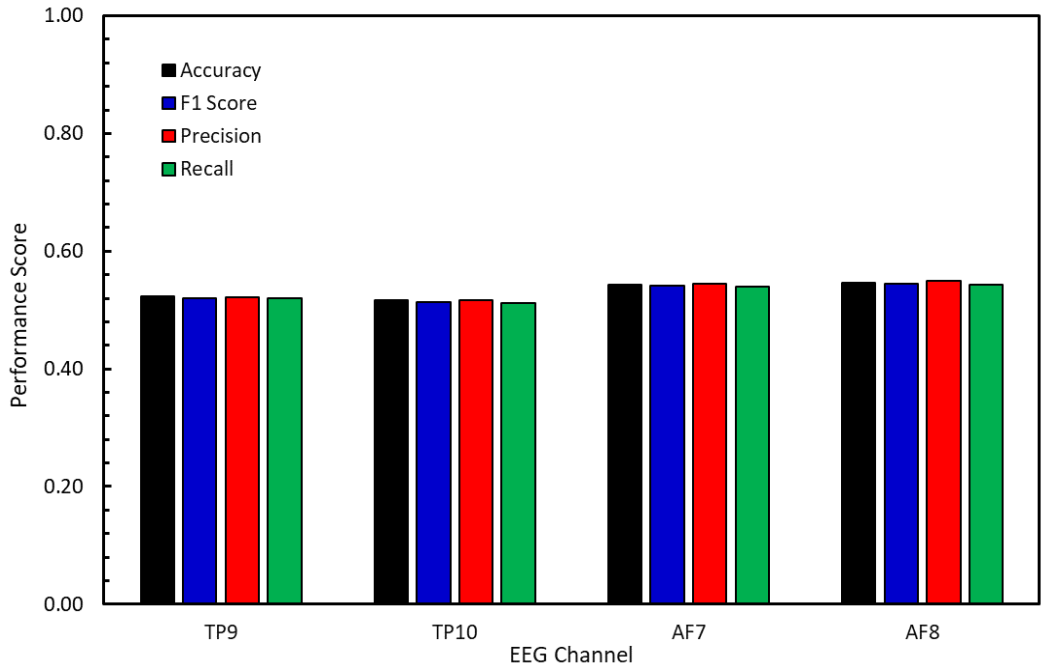

**Figure 4** Machine Learning Algorithm Evaluations on singular EEGs. The AF8 EEG channel performs at the highest accuracy, but all the EEG channels seem to all be performing at a low accuracy percentage

To see the more in-depth strengths of classification for each STEM task, each EEG channel was also evaluated for their accuracies on each task. Overall, from the chart of the results seen in Figure 5, it is seen that the AF8 EEG channel performed the best on TOL and MathProc, AF7 performed the best on BCST and MSPAN, and TP9 performed the best on connection.

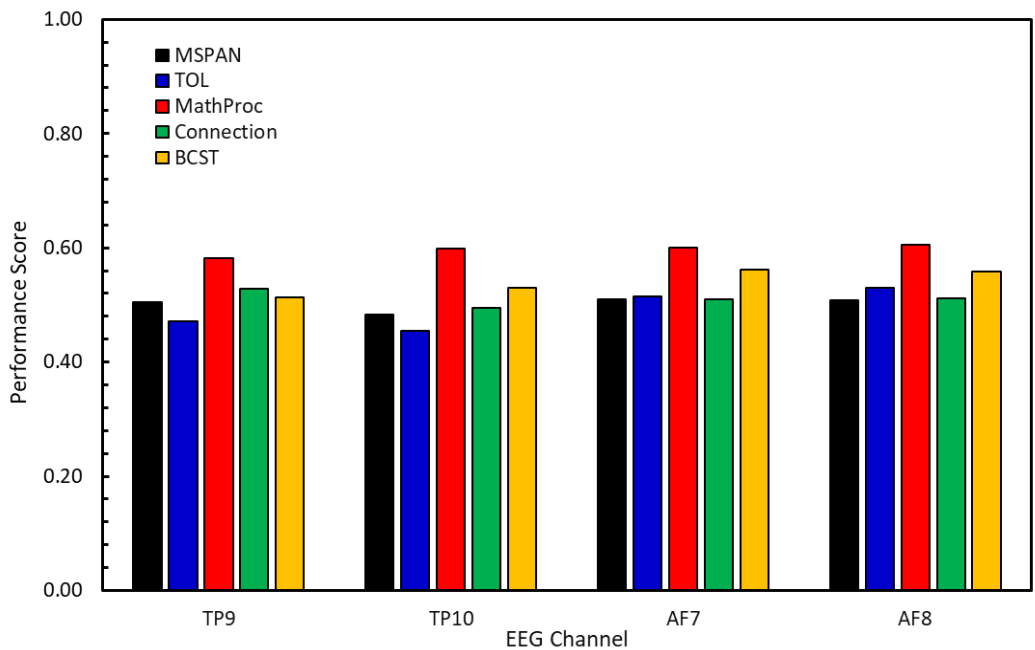

**Figure 5** Classification Label Evaluation on Singular EEGs. The AF8 channel has the highest accuracy of TOL and MathProc, AF7 has the highest accuracy for BCST and MSPAN, TP9 has the highest accuracy on connection, and TP10 has the lowest accuracy for everything

## 3.3 Power Spectral Densities of STEM Activities

Instead of just relying on the results of the machine learning algorithm accuracies, a visualization of the averages of each frequency band and task was taken for each task. In Figure 6, there seem to be good margins based on the power spectral densities of different tasks. The delta frequency band seems to provide the most information when trying to differentiate between different STEM tasks. It is also seen that in the theta and alpha power spectral densities, the math problem solving task usually has the lowest power spectral density while the Wisconsin card sort holds higher power spectral density for the most part. In the beta and gamma waves, the Wisconsin card sort activity usually had the highest power spectral densities when comparing each EEG channel individually with one another.

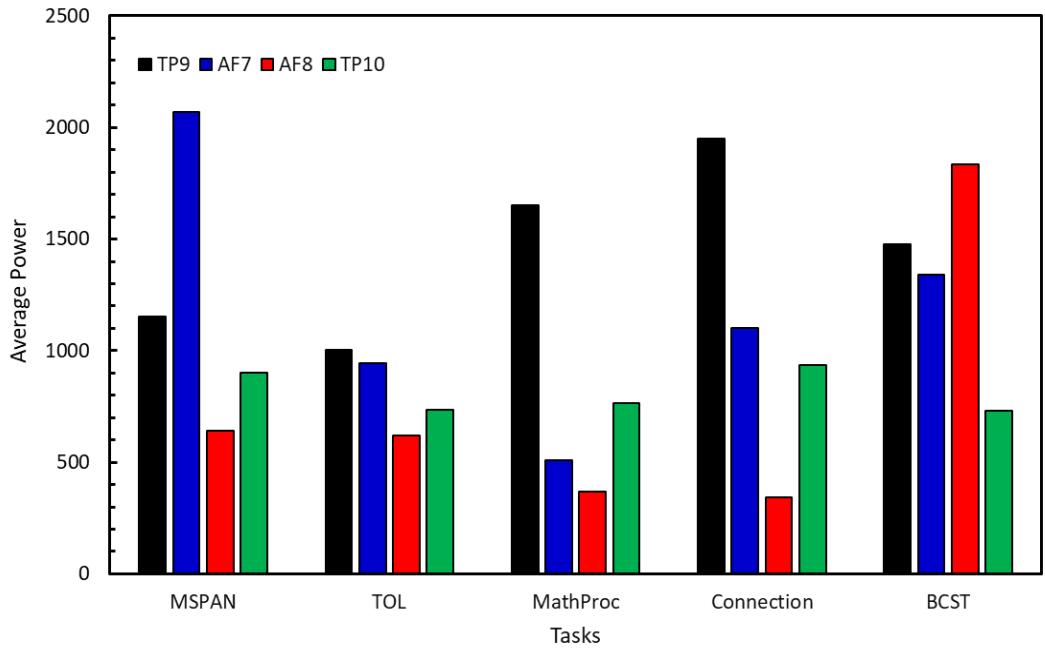

(a) Delta

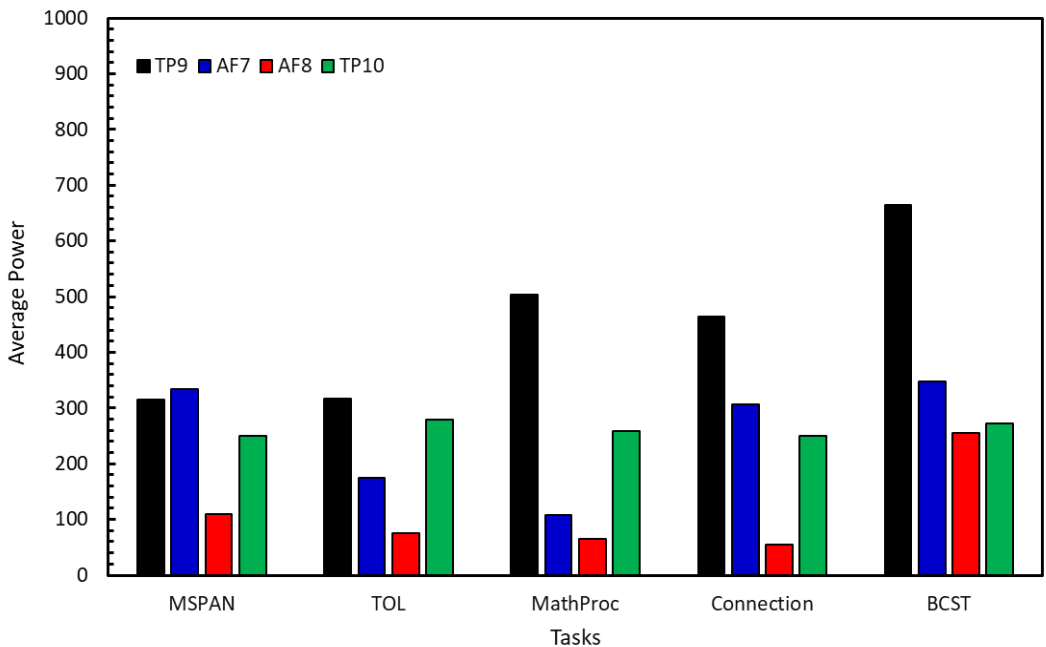

(b) Theta

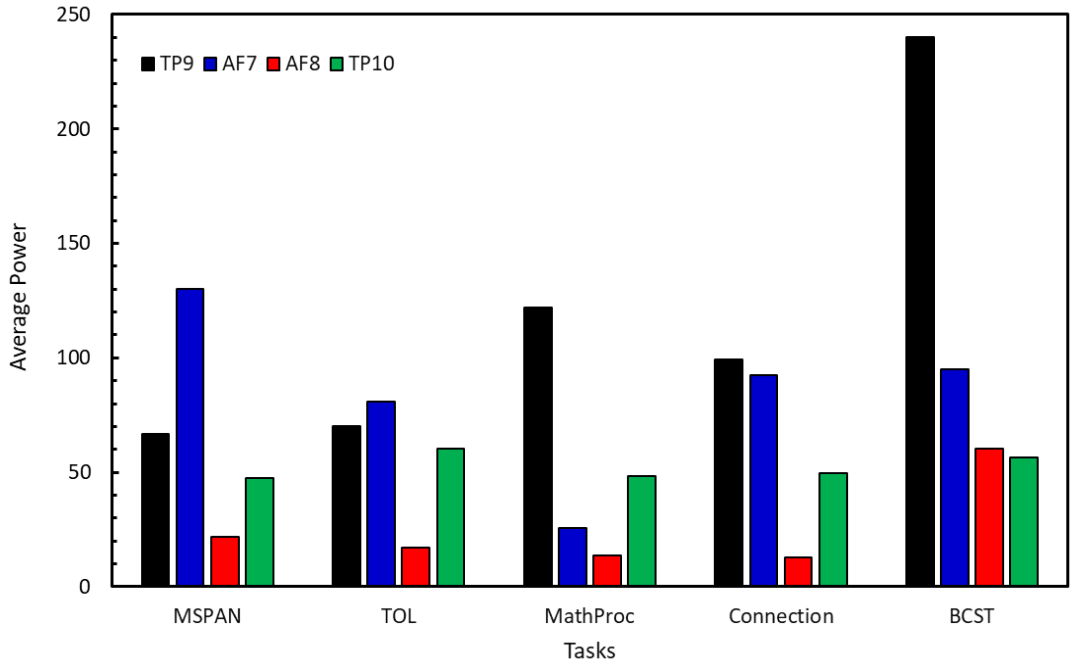

(c) Alpha

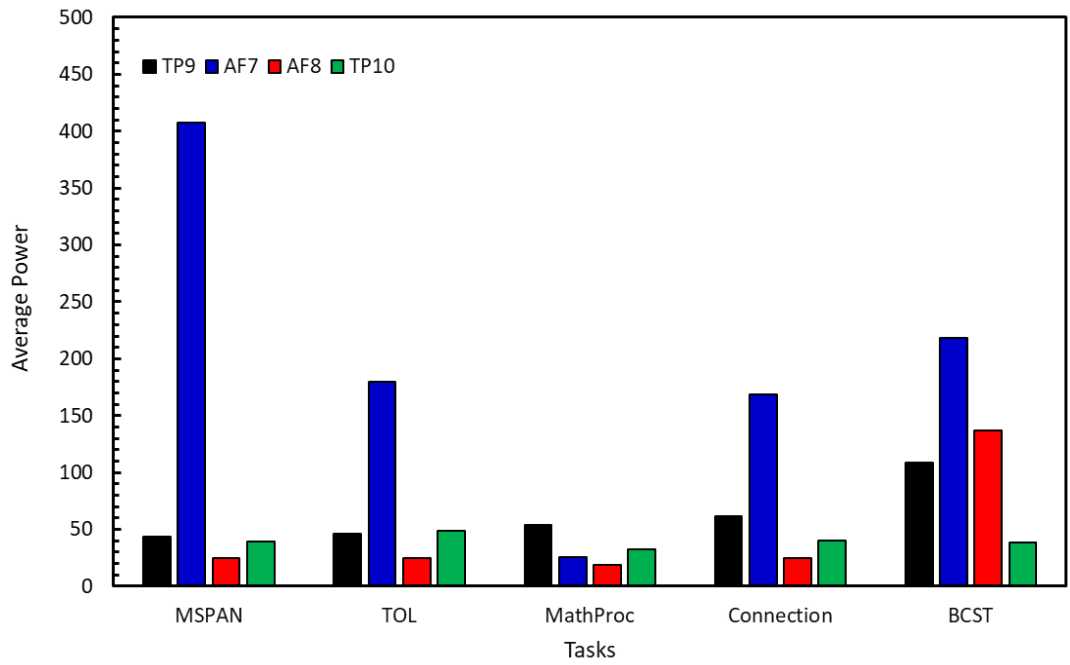

(d) Beta

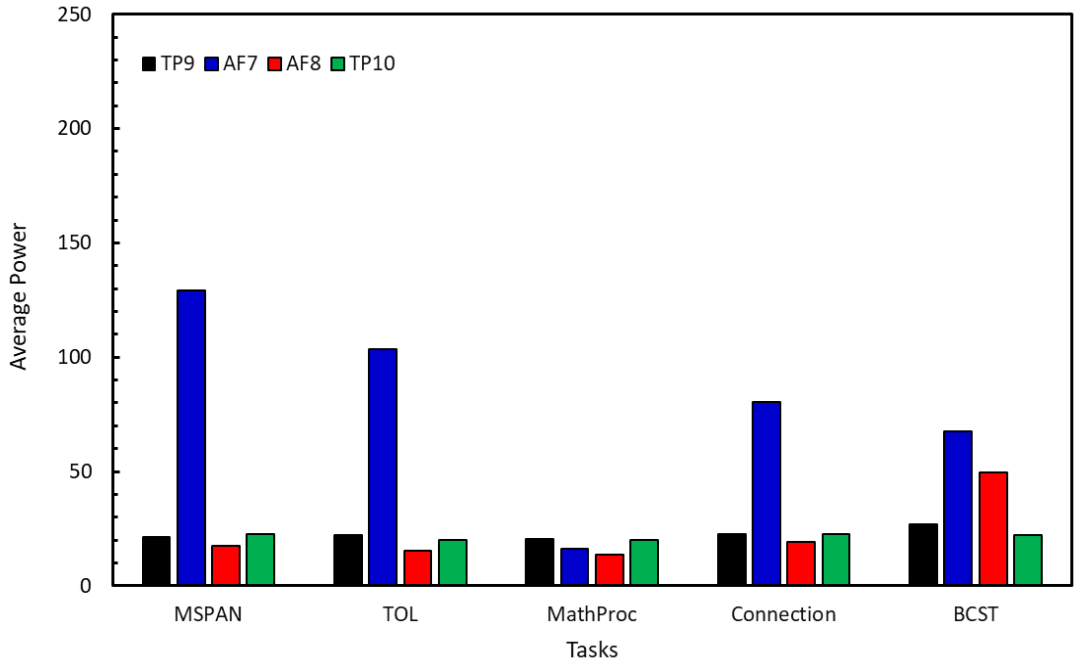

(e) Gamma

**Figure 6** Power Spectral Density Averages on Four EEGs. The BCST activity has the highest power spectral densities in the theta and alpha frequency waves

  The overall power band power spectral densities of all the EEG sensor locations were also taken, with Figure 7 representing the results. It is seen that BCST exhibits the highest power spectral density throughout the brain while MathProc had the lowest power. In general, all the tasks seem to be very distinct from each other, meaning that the various STEM tasks are very distinct from one another.

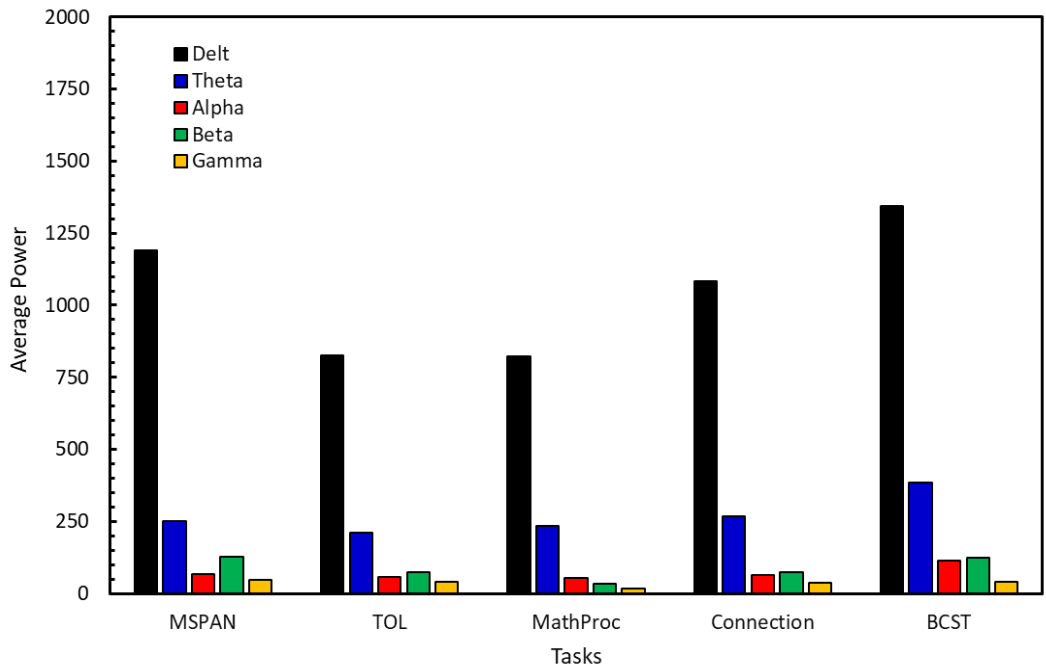

**Figure 7** Overall Power Spectral Density Averages. The delta frequency wave has the biggest fluctuation between the different activities

## 4. Discussion

In this study, the best machine learning algorithm for decision tree-based algorithms was found, the tasks were further analyzed for patterns, each brain region was analyzed for its importance, and correlations between brain region and tasks were discovered. It is noticed that the Random Forest performs at the highest level during all interval sizes compared to the XGBoost and Bagging Classifier. Then, the XGBoost is found to have the lowest performance compared to both the XBoost and Random Forest most of the time. This creates an understanding that a more uniform distribution of decision trees, as seen from the Random Forest, is the best decision tree machine learning algorithm for STEM Learning Classifications from EEGs. However, parallel tree distributions, as seen from the XGBoost, may not be as effective due to their lower performance overall.

From this created algorithm, the different asks were also compared for accuracy. It was found that cognitive flexibility was the most clearly seen. Several studies in the past agree with these results, especially with the result of the Random Forest performing well under EEG data. For example, Kumari et. al found that during emotion analysis of the brain, the Random Forest created a classification result of 98.12%, performing the best when compared with other machine learning algorithms(Kumari et al., 2022). However, in Parui et. al's study, they found that the XGBoost performed the best compared to other algorithms (Parui et al., 2019). These differences in what the best algorithm is could be due to different characteristics throughout the data during human emotion classification from EEG data. XGBoost may have performed the best in Parui's study due to including data with less distinct differences between classification labels while Kumari's study may have had data that was more easily differentiable. Since there weren't any relevant literature comparing Random Forest and XGBoost specifically on STEM Learning tasks, this discovery could prove to be beneficial in future research. Since Random Forest established itself as the highest-performing algorithm, it was then used in all the other brain region analysis methods in this study. Random Forest most likely performed at the highest level of accuracy due to its more simplistic nature of training compared to the XGBoost that relies on gradient boosting. This means that brain data between different human behaviors is can be more directly seen so a complex classification approach may be unnecessary compared to the Random Forest Classifier (Bentéjac et al., 2021).

Delving into the power spectral densities of STEM activities furthered our understanding, revealing distinct patterns for each task, particularly in the delta frequency band. With the delta frequency range showing the most fluctuations in terms of power spectral density between the different tasks comprehensively, it seems to stand out as the most important biomarker when analyzing power spectral densities. While this may seem strange since the delta wave is known more for a relaxed state of mind, this vast variation in power spectral densities between the different cognitive tasks could point to different levels of overall concentration (Harmony, 2013). It is also concluded that in general, cognitive flexibility seemed to take up more student concentration while arithmetical functioning is not as difficult as the higher frequency waves such as Theta, Alpha, Beta, and Gamma are usually the highest in cognitive flexibility but lowest in math problem solving. When looking at previous research, several agree that delta power spectral densities are a major marker in analysis to better understand cognitive tasks. The study found that delta oscillations increase during activities such as mental calculations (Harmony, 2013). It was also found that delta power increases were seen in the frontal region during motor response activities and semantic tasks, showing that increases in delta frequencies in mental tasks are connected to sensory differences. This could explain why the machine learning algorithms in this study classified the STEM learning tasks at such a high accuracy because of the constant external signals and information required for each subject to solve each STEM learning task.

To find correlations between each brain region and each of the different STEM learning tasks, the same models used to calculate each brain region's level of importance were used, and then we calculated several evaluation metrics for each STEM learning task label. The results revealed that distinct brain regions play pivotal roles in the cognitive processes underpinning STEM learning. The frontal lobe showed to be most important compared to the temporoparietal regions, with the Right Frontal lobe correlating with mathematical processing and planning, while the Left Frontal lobe connects the most to cognitive flexibility and mental flexibility. The Left Temporoparietal lobe distinguishes itself in tasks requiring cognitive connections. However, it is worth noting that the differences in classification accuracy between the different brain regions are very minimal, so all the neural processes are interconnected for all the STEM Learning tasks, it is just that certain brain regions centralize on these topics a little more than the others. Previous research agrees with these findings, with Collins et. al finding from a computational model of a human that the frontal lobe represents reasoning, decision-making, and adaptive behavior (Collins & Koechlin, 2012). This connects to our findings that the left frontal lobe and right frontal lobe represent mathematical processing, planning, cognitive flexibility, and mental flexibility. However, our findings show that different sides of the brain, left or right, portray slightly different strengths in different STEM learning activities. In Spierer et. al's study, it was found that the Temporoparietal Junction serves as a part of the brain that helps the perception of the timing of sensory events and higher-order cognitive functions based on stimuli (Spierer et al., 2009). This connects to our findings of the left temporoparietal lobe representing cognitive connections.

To determine each brain region's level of importance overall in STEM Learning task classification, each EEG sensor was used again using the Random Forest, and several evaluations of its performance were recorded. Overall for all the STEM Learning tasks, the right frontal lobe seemed to perform at the highest level, followed by the left frontal lobe, left temporoparietal lobe, and then the right temporoparietal lobe. However, the differences in accuracy between the different lobes were very miniscule, different only be a couple percent. This leads to the finding that different STEM Learning tasks can be detected in different brain regions due to interconnected neural networks, stronger suggesting connections between the frontal and temporoparietal lobe to be the main contributors. This is further supported by how each lobe when independently predicting different STEM learning tasks performed at the 50% range, but when all the lobes are used together, it reached the 90% range. Nielsen et. al supports the fact that there are numerous interconnected neural networks between brain regions, explaining why different brain regions still can detect various STEM activities (Nielsen et al., 2013).

### 4.1 Real-Life Applications

Exams and quizzes may not fully capture a student's real-time understanding and learning progress. These assessments often focus on whether or not the student can correctly answer each question instead of the process of thinking that leads to the answer. Noninvasive electroencephalogram (EEG) sensors can actively monitor student brain waves during class, being able to identify the learning patterns students go through to learn. This information can be used to determine the best way for a student to learn and adapt teaching methods accordingly. EEG sensors, when combined with machine learning, can allow teachers to adjust their lectures based on the preferred learning methods of most students. This is a more effective approach than relying solely on teacher observations and tests, which may not fully capture a student's learning performance.

### 4.2 Limitations

This study used brain data from participants who are mostly university students. While the data sample used was relatively large, the brain processes of university students could be different from other age groups. In the future, a wider age group range of subjects should be used to analyze possible variations in brain processes between different ages.

### 5 Conclusions

This study conducted a comprehensive analysis of machine learning algorithms for STEM learning classifications using EEG data. Random Forest was identified as the most effective algorithm, outperforming XGBoost and Bagging Classifier. Notably, the delta frequency band emerged as a crucial biomarker for distinguishing between STEM tasks, highlighting the most differences in power spectral densities between different tasks. Cognitive flexibility, which is the STEM activity of Wisconsin Card Sort, was found to require greater student concentration due to overall higher power spectral densities in higher ranges of frequencies while math problem solving took the least focus. The study also highlighted the importance of different brain regions in STEM learning, with the right frontal lobe excelling in mathematical processing, the left frontal lobe in cognitive flexibility, and the left temporoparietal lobe in cognitive connections. These findings align with previous research and suggest that distinct brain regions play significant roles in STEM learning, although the differences in classification accuracy between regions were minimal, highlighting their interconnected nature.

# 6 Acknowledgements

This study was partially supported by 2023 UFA funding at NIU.

# 7 Author contributions

Conceptualization: Jaejin Hwang, Ryan Cho; Methodology: Mobasshira Zaman, Ryan Cho; Formal analysis and investigation: Ryan Cho; Writing - original draft preparation: Ryan Cho; Writing - review and editing: Jaejin Hwang, Kyu Taek Cho, Ryan Cho; Funding acquisition: Jaejin Hwang, Kyu Taek Cho; Resources: Jaejin Hwang; Supervision: Jaejin Hwang.

# 8 Funding

This study was partially supported by 2023 UFA funding at NIU.

# 9 Competing interests

The authors declare that they have no financial or non-financial interests that are directly or indirectly related to the work submitted for publication.